\documentclass[showkeys, pra,aps,twocolumn,twoside,showpacs, superscriptaddress,longbibliography]{revtex4-2}
\usepackage[T1]{fontenc}
\usepackage{amsfonts}
\usepackage{amssymb}
\usepackage{mathrsfs}
\usepackage[intlimits]{amsmath}
\usepackage{bm, bbm}
\usepackage[dvips]{graphicx}

\usepackage[T1]{fontenc}
\usepackage[utf8]{inputenc}
\usepackage {hyperref}
\usepackage{amsmath}
\usepackage{latexsym}
\usepackage{soul,xcolor}
\usepackage{color}
\usepackage{graphicx}
\usepackage[export]{adjustbox}
\usepackage{subcaption}

\usepackage{physics}
\usepackage{appendix}

\def\be{\begin{equation}}
\def\ee{\end{equation}}
\def\ben{\begin{eqnarray}}
\def\een{\end{eqnarray}}

\begin{document}

\title{Decoherence from the light bending interaction}

\author{T. Bazylewicz} 
\affiliation{Center for Theoretical Physics, Polish Academy of Sciences, 02-668 Warsaw,  Poland}
\affiliation{Faculty of Physics,  University of Warsaw, 02-093 Warsaw, Poland}

\author{M. Szczepanik} 
\affiliation{Faculty of Applied Physics and Mathematics, Gdańsk University of Technology,  80-233 Gdańsk, Poland}

\author{J. K\l os}
\affiliation{Faculty of Physics,  University of Warsaw, 02-093 Warsaw, Poland}

\author{J. K. Korbicz}
\affiliation{Center for Theoretical Physics, Polish Academy of Sciences, 02-668 Warsaw,  Poland}

\begin{abstract}
    We analyse a decoherence effect, caused by the gravitational interaction between a massive body and the electromagnetic field. Assuming a quantum version of the light bending interaction, we show that it leads to decoherence  of the mass if the light is not observed. Using the extreme weakness of the gravitational coupling, we derive explicitly the decoherence lengthscales for general states of the central mass and for both thermal and coherent light. Predictably, the effect is very faint for anything but hugely energetic light, however from the fundamental point of view of co-existence of both gravitation and quantum theories, it is there.  Since effectively the studied system is a quantum optomechanical system, we hope our results, properly rescaled, will be also useful in optomechanics.   
\end{abstract}

\maketitle

\section{Introduction}

Recent years have witnessed an increased interest in the possible interplays between quantum mechanics and gravitation, motivated in part by the recent advances in our understanding of the former brought by the quantum information science \cite{Bose_spin_entanglement}, \cite{Marletto_gravity_ent}, \cite{Marletto_quantum_inf} and in part by a spectacular advancement of experimental techniques \cite{Westphal_measuremnt}, to name just a few. Widely discussed questions are if gravitational interaction can lead to quantum entanglement and to decoherence. The idea that gravitational self-interaction may provide a decohering, and thus a classicization, mechanism dates back to the seminal works of Di\'osi \cite{Diosi_master} and Penrose \cite{penrose_gravity}. More recently, another ideas have been proposed, based on a gravitational interaction between different degrees of freedom, in particular the, so called,  universal gravitational decoherence \cite{Pikovski2015}.

In this work, we study yet another gravity-mediated decoherence effect, induced by the interaction of a mass with light, i.e. by the light bending interaction \cite{Bodenner2003}. It is postulated to hold also in the quantum regime, keeping however the gravitational field classical, and the consequences are studied. This system was proposed and analysed in \cite{Light_Matter_H}, where however the main focus was on the possible entanglement generation. It was shown via effective loss of purity of the gravitating mass, signifying decoherence. However, the decoherence itself was not analysed, in particular the relevant lengthscales were not derived. We complement and generalize those studies here by: i) deriving explicit expressions for decoherence lengthscales in the model; ii) generalize to the arbitrary, mixed states of the gravitating mass; iii) consider both coherent and thermal light. Although the effects are expectedly faint for anything but astronomical bodies, they are there nevertheless, providing another example of a  gravitation-driven decoherence mechanism (provided of course that the proposed quantization of the light bending interaction works). Our results have also relevance for the field of quantum  optomechanics \cite{Knight, Aspelmeyer2014} since the resulting interaction has the same shape as in the optomechanical studies. With the appropriate change of the coupling constants, our formulas provide explicit decoherence lengths for optomechanical systems in the regime of a weak interaction but for a general state of the resonator.

\section{Effective dynamics}
We will use the results of \cite{Light_Matter_H}, where it was derived the effective light bending interaction between the massive harmonic oscillator of mass $m$ and frequency $\Omega$ and the electromagnetic field, localized at the mean distance $r$ from the oscillator: 
\begin{equation}\label{V}
    V_{k\nu}=-g_0(b^{\dagger}+b)\otimes \hbar \omega_k a^{\dagger}_{k\nu}a_{k\nu},
\end{equation}
where $b$, $b^\dagger$ are the annihilation and creation operators of the central oscillator, $a^{\dagger}_{k\nu}$, $a_{k\nu}$ of the field mode with momentum $k$ and polarization $\nu$,  and 
\begin{eqnarray}\label{g}
    g_0 =\frac{4GM}{r^2c^2}\sqrt{\frac{\hbar}{2M\Omega}} =\delta\varphi\frac{\delta x_{zpf}}{r}
\end{eqnarray}
is the effective, dimensionless coupling strength. It is a product of the classical angle of the light deflection at the distance $r$, $\delta\varphi=4GM/(rc^2)$ \cite{Bodenner2003}, with $G$ being the gravitational constant, and the zero point fluctuations of the oscillator $\delta x_{zpf}$, referenced to the distance $r$. There are several approximations assumed in deriving \eqref{V}, see \cite{Light_Matter_H}, the main being that:  i) the mass is large enough not to feel the decoy of the light; ii) its oscillations are small compared to the distance $r$ and iii) we keep only the linear term, which is subsequently quantized; iv) the light scattering is elastic and no change of the polarization happens, i.e. $k$,$\nu$ parameters are the same before and after the scattering.

The total Hamiltonian is thus given by:
\begin{equation} \label{Htot}
    H = H_S+H_E+V = \Omega b^\dagger b + H_E - g_0(b^\dagger + b)\otimes H_E.
\end{equation}
where  we assumed that in general a collection of modes is present:
\begin{equation}
    H_E = \sum_{k\nu} \hbar\omega_ka^\dagger_{k\nu}a_{k\nu}.
\end{equation}
As already noted in \cite{Light_Matter_H}, this is nothing else but a Hamiltonian of the well known optomechanical coupling between light and massive oscillator; see e.g. \cite{Aspelmeyer2014} for a review. We treat the oscillator as the system of interest $S$ while the light constitutes its environment $E$.

The evolution, generated by \eqref{Htot} can be solved exactly by e.g. changing to the interaction picture and using the Magnus expansion, which in this case has only two non-zero terms \cite{Knight}. The resulting evolution is given by the product (in what follows, we will use the units $\hbar=c=1$):
\begin{equation}\label{U}
    U_{int}^{I}(t)=U_{SE}(t)\left[{\bf 1}_S\otimes\Phi(t)\right],
\end{equation}
where:
\begin{eqnarray}
     &&  U_{SE}(t) = \exp\left\{\left(\lambda(t)b^\dagger-\overline{\lambda}(t)b\right)\otimes H_E\right\}, \label{USE}\\
     &&  \lambda(t) = \frac{g_0}{\Omega}(e^{i\Omega t}-1).
\end{eqnarray}
is the entangling evolution between the oscillator with the light and:
\begin{align}\label{PHI}
     \Phi(t)=\exp\left\{-ig_0^2\left[\Omega t - \sin(\Omega t)\right] (H_E/\Omega)^2\right\},
\end{align}
is a "phase operator", acting on the light part only. We emphasize that the above solution \eqref{U}-\eqref{PHI} is exact, there were no perturbative approximations used in deriving it from \eqref{Htot}


\section{General reduced dynamics of the oscillator}
The expression for the reduced state of the system has been known in the literature for many years
\cite{Knight, Aspelmeyer2014} but mostly for specific initial states of the system, e.g. for the pure coherent state. Here we derive a general formula for $\rho_S(t)=\Tr\rho_{SE}(t)$, valid for any initial state of the system $\rho_S$, and calculate the position basis matrix elements $\rho_t(x,x')=\langle x|\rho_S(t)|x'\rangle$.

To solve for the dynamics, we make the usual open system assumption that the total initial state is uncorrelated: $\rho_{SE}(0)=\rho_S\otimes\rho_E$. Although it  simplifies the calculation significantly, this assumption must be treated with care here as the gravitational interaction between the mass and the light in principle exists always as the potential is long range $V \sim 1/r$ \cite{Bodenner2003, Light_Matter_H}. However, one could imagine  a situation where the mass and the light are first separated far enough for the interaction to be neglected and then brought close enough to each other in an adiabatic manner. Using the product assumption and decomposing the operator $U_{SE}(t)$ of \eqref{USE} in the Fock basis of the light and changing back to the Schroedinger picture, we obtain the solution of the total dynamics:
\begin{align}
    & \rho_{SE}(t)=\sum_{EE'}e^{-iH_St}\hat{D}\left(E\lambda(t)\right)\rho_S \hat{D}\left(E'\lambda(t)\right)^{\dagger}e^{iH_St}\otimes\nonumber\\
    & \bra{E}\rho_{E}\ket{E'}e^{-iH_Et}\Phi(t)\ket{E}\bra{E'}\Phi^{\dagger}(t)e^{iH_Et}. \label{rhoSE}
\end{align}
Here $\hat D(\alpha)$ is the optical displacement operator and $\ket{E}$ is the energy eigenstate of the light:
\begin{align}
    \ket{E}=\bigotimes_{k\nu} \ket{n_{k\nu}},
\end{align}
corresponding to the total energy $E=\sum_{k\nu}\omega_kn_{k\nu}$.

Assuming that the light is not under our control, we trace its degrees of freedom from the total state \eqref{rhoSE} and calculate $\rho_S=\Tr_E{\rho_{SE}}$. Due to the cyclic property of the trace and the fact that $[\Phi(t),H_E]=0$, we obtain a term: 
\begin{eqnarray}
    \Tr{e^{-iH_Et}\Phi(t)\ket{E}\bra{E'}\Phi^{\dagger}(t)e^{iH_Et}}=\delta_{EE'},
\end{eqnarray}
which leads to the following expression of the state of the system:
\begin{eqnarray}\label{generaldensitymatix}
    &&\rho_S(t)=\nonumber\\
    &&\sum_E p(E) e^{-i\hat{H}_St}\hat{D}\left(E\lambda(t)\right)\rho_S\hat{D}^{\dagger}\left(E\lambda(t)\right) e^{i\hat{H}_St},\label{rhoSE2}
\end{eqnarray}
where $p(E)=\langle E |\rho_E| E\rangle$ is the initial distribution of energy in the light environment.

To proceed, it is convenient to represent $\rho_S$ using its $P$-representation
\cite{Sudarshan}\cite{Gaubler}
, i.e. representation diagonal in the coherent state basis $\ket\alpha$, which is known to always exist for wide enough class of distributions $P(\alpha)$ (see e.g. \cite{Perelomov1986}):
\begin{align}
   \rho_S = \int \frac{d^2\alpha}{\pi} P(\alpha)\ket{\alpha}\bra{\alpha}. 
\end{align} 
This simplifies the displacement operation since $D(\beta)\ket\alpha\bra\alpha D(\beta)^\dagger = \ket{\alpha+\beta}\bra{\alpha+\beta}$. Further using that $e^{-i\hat{H}_St}\ket{\alpha}=\ket{e^{-i\Omega t}\alpha}$, we obtain:
\begin{eqnarray}\label{matrixinPrep}
    &&\rho_S(t) = \nonumber\\
    &&\sum_E p(E)\int \frac{d^2\alpha}{\pi}P(\alpha)e^{-i\hat{H}_St}\ket{E\lambda(t)+\alpha}\bra{E\lambda(t)+\alpha}e^{i\hat{H}_St}\nonumber\\
    &&=\sum_E p(E)\rho_S(t; E),
    \label{rhoS}
\end{eqnarray}
where  
\begin{align}
    \rho_S(t; E)=\int \frac{d^2\alpha}{\pi}P(e^{-i\Omega t}\alpha)\ket{E\lambda'(t)+\alpha}\bra{E\lambda'(t)+\alpha}
\end{align}
is a contribution to $\rho_S(t)$ from the field with a definite energy $E$ and $\lambda'(t)=e^{-i\Omega t}\lambda(t)$. 
In the last step, we have moved the time dependence generated by $H_S$ from the coherent states to the  $P$-representation for the sake of simplicity. 

In what follows, we will be interested in the decoherence effects, described by \eqref{rhoS}. Since the interaction and the self-Hamiltonian $H_S$ do not commute, it is a non-trivial question what is the chosen basis of the pointer states \cite{Zurek1993} ,\cite{Schlosshauer2007}, \cite{singh_korbicz2024}. But since the gravitational interaction is very weak, dominated by the self-Hamiltonian, one expects that the approximate pointer basis will be close to the energy eigenstates of the oscillator, cf. \cite{Cucchietti2007}, \cite{PazZurek1999}. Indeed, in \cite{Light_Matter_H} the loss of purity of $\rho_S(t)$ was calculated in the energy basis for a coherent initial state. One could in principle deduce decoherence factors from that. However, here we take another approach and look at the decoherence effects in the position basis, which is the basis singled out by the gravitational coupling \eqref{V}. The motivation is to see to what extent the gravity-mediated interaction reduces coherences in the real space. Let us denote by: 
\begin{align}
  &\rho_t(x,x';E) = \langle x| \rho_S(t; E)| x'\rangle=\\
  &\int \frac{d^2\alpha}{\pi}P(e^{-i\Omega t}\alpha)\bra x\ket{E\lambda'(t)+\alpha}\bra{E\lambda'(t)+\alpha}\ket{x'} 
\end{align}
matrix elements of the fixed energy component of the state \eqref{rhoS}. We note that the $E=0$ component corresponds to the matrix elements of the freely evolving central oscillator, cf. \eqref{generaldensitymatix}. Using the well-known formula for the wave function of the coherent states:
\begin{align}
   \langle x|\alpha_1+i\alpha_2\rangle=\left(\frac{2\kappa}{\pi}\right)^{\frac{1}{4}}\exp\left[-\left(\xi-\alpha_1\right)^2+2i \alpha_2 \xi  -i\alpha_1\alpha_2\right], 
\end{align}
where $\alpha=\alpha_1+i\alpha_2$ and $\xi = \sqrt{\mathcal{\kappa}}x$ with $\kappa = M\Omega/(2\hbar)$ is the dimensionless position,  we obtain:
\begin{eqnarray}\label{rhoinprep}
        &&\rho_t(x,x';E)= \sqrt{\frac{2\kappa}{\pi^3}} \int d^2\alpha P(e^{-i\Omega t}\alpha) \times\nonumber\\
        &&\exp\Big\{ \left(\xi-\alpha_1-E\gamma_1(t)\right)^2+\left(\xi'-\alpha_1-E\gamma_1(t)\right)^2 \\
        &&+2i\Delta \left(\alpha_2+E\gamma_2(t)\right)\Big\}\nonumber, 
\end{eqnarray}
where $\Delta=\xi-\xi'$ is the relative, dimensionless distance and
\begin{eqnarray}
   &&\gamma_1(t)=\frac{g_0 }{\Omega}\left[1-\cos(\Omega t)\right],\label{gamma1}\\
    &&\gamma_2(t)=\frac{g_0 }{\Omega}\sin(\Omega t).\label{gamma2}
\end{eqnarray}

The above expression can be further simplified by passing to the symmetric characteristic function $\chi(\beta)$ of the initial state, defined as  $\chi(\beta)= \Tr[D(\beta)\rho_S]$ and related to the $P$-representation via \cite{Perelomov1986} :
\begin{align}
P(\alpha)= \int \frac{d^2\beta}{\pi} \chi(\beta)e^{|\beta|^2/2} e^{-\beta\overline\alpha+\overline\beta\alpha}    
\end{align}
After some algebraic manipulations and observing the fact that the integral w.r.t. $\alpha_2$ in \eqref{rhoinprep} produces a delta function $\pi \delta(\beta_1+\Delta)$, we finally obtain:
\begin{eqnarray}\label{generalform}
    &&\rho_t(x,x')= \sqrt{\frac{\kappa}{\pi^2}}\int dp\chi_t(-\Delta,p)e^{-i\xi_+p} \nonumber\\
    && \times\sum_E p(E)\exp\left\{2iE\left[\gamma_1(t) p +\gamma_2(t) \Delta\right]\right\},
\end{eqnarray}

where we introduced $\xi_+=\xi+\xi'$ and:
\begin{eqnarray}
    \chi_t(\beta_1,\beta_2)= \chi(e^{-i\Omega t}\beta),
\end{eqnarray}
and renamed the integration variable $\beta_1$ as $p$.  Eq. \ref{generalform} is the most general formula for the position-basis matrix elements of the reduced state. It has the following form:
\begin{align}\label{rhofinal}
    \rho_t(x,x')=\int dp \rho_{0t}(x,x';p) F_t(p; \Delta), 
\end{align}
where $\rho_{0t}(x,x';p)=\sqrt{\kappa/\pi^2}\chi_t(-\Delta,p)e^{-i\xi_+p}$ are the $p$-components of the freely evolving, i.e. $E=0$, matrix elements. Here, $F_t(p; \Delta)$ is the influence function, that encompasses both the mechanical influence of the environment on the oscillator and the position decoherence effects:
\begin{align}
    & F_t(p; \Delta)=\sum_E p(E)\exp\left\{2iE\left[\gamma_1(t) p +\gamma_2(t) \Delta\right]\right\}\label{F}\\
    & = \sum_{\{n_{k\nu}\}} p\left(\{n_{k\nu}\}\right) \prod_{k,\nu} \exp\left\{2i\omega_kn_{k\nu}\left[\gamma_1(t) p +\gamma_2(t) \Delta\right]\right\},
\end{align}
where $\sum_{\{n_{k\nu}\}}$ is the sum over all possible occupation numbers of the field (we recall that $E=\sum_{k\nu}\omega_kn_{k\nu}$ ).  We note that this function gives no contribution for $\Delta=0$. Mathematically, it is a discrete Fourier transform of the initial energy distribution of the light $p(E)$. Below we calculate it for the case of a thermal and coherent light.


\section{Decoherence from a thermal light}
For thermal light $\rho_{E}=e^{-\beta H_E}/Z$, where $\beta=1/(k_BT)$,  we have:
\begin{align}
 p(\{n_{k\nu}\})=\prod_{k,\nu} e^{-\beta \omega_k n_{k\nu}}(1-e^{-\beta\omega_k}).   
\end{align}

Then $F_t(p; \Delta)$ can be easily calculated through the geometric series formula:
\begin{eqnarray}\label{Fczynnik}
    F_t(p; \Delta)=\prod_{k,\nu}\frac{e^{\beta\omega_k}-1}{e^{\beta\omega_k}-\exp\left[2i\omega_k(\gamma_1(t)p +\gamma_2(t)\Delta) \right]}
\end{eqnarray}
The next standard step would be now passing to the logarithm $\ln F_t(p; \Delta)$ and then to the continuum limit with some spectral density $J(\omega)$ in order to change the problematic product over the modes into an integral. However, due to the complicated functional form of \eqref{Fczynnik}, the resulting integral over the modes presents little simplification of the problem. Instead of that, for the purpose of the current work we will approximate \eqref{Fczynnik} using the fact that the effective coupling strength $g_0$ of \eqref{g}, present in $\gamma_{1,2}(t)$, cf. \eqref{gamma1}, \eqref{gamma2} is for realistic scenarios extremely small \cite{Light_Matter_H}. To be precise, we assume that: 
\begin{align}\label{approx}
    g_0\frac{\omega_k}{\Omega} \ll 1, 
\end{align}
as this is how $g_0$ enters \eqref{Fczynnik}, cf. \eqref{gamma1}, \eqref{gamma2}.


\subsection{High temperature approximation }
We expect that due to the extremely weak nature of the gravitational coupling, decoherence effect will manifest itself when the light is highly energetic. Therefore, on top of the condition \eqref{approx}, let us further assume  the high temperature regime, i.e. $\hbar\omega_k\beta \ll 1$. This also allows to simplify the calculations. We obtain from \eqref{Fczynnik} that $\omega_k$ drops out and the product is $k$-independent:
\begin{align}
    & F_t(p; \Delta) \approx \prod_{k,\nu}\frac{\beta}{\beta-2i(\gamma_1(t)p +\gamma_2(t)\Delta)}\\
    & = \frac{1}{\left[1-\frac{2i}{\beta}(\gamma_1(t)p+\Delta\gamma_2(t))\right]^{2N}}\\
    & \approx 1+\frac{4iN}{\beta}\left[\gamma_1(t)p+\Delta\gamma_2(t)\right]\nonumber\\
    & - \frac{4N(N+1)}{\beta^2}\left[\gamma_1(t)p+\Delta\gamma_2(t)\right]^2
\end{align}
where $N$ is the number of modes and in the second step of approximation we assumed that $g_0$ is small enough so that $g_0/(\beta\hbar\Omega)=g_0k_BT/\hbar\Omega \ll 1$ also holds. We kept the orders up to quadratic since we will be interested in $|\rho_t(x,x')|^2$ rather then $\rho_t(x,x')$ as we are interested in the magnitude of the decay of the spatial correlations. We note that in this approximation, the impact of the environment does not depend on the frequency of  the light, it dropped out after the first step. Substituting the above approximation into \eqref{rhofinal}, we obtain:
\begin{eqnarray}
    &&\rho(x,x')\propto\left[ 1 +4iN\Delta \frac{\gamma_2(t)}{\beta} - 4N(2N+1)\Delta^2\frac{\gamma_2^2(t)}{\beta^2}\right]\rho_0-\nonumber\\
    && \left[4N\frac{\gamma_1(t)}{\beta}+8iN(2N+1)\Delta\frac{\gamma_1(t)\gamma_2(t)}{\beta^2}\right]\partial\rho_0+\nonumber\\
    && 4N(2N+1)\frac{\gamma^2_1(t)}{\beta^2}\partial^2\rho_0
\end{eqnarray}
Above we introduced a simplified notation, where $\rho_0 = \int dp \chi_t(-\Delta,p)e^{-i\xi_+p}$ denotes the matrix elements of the freely evolving oscillator and $\partial\rho_0=\partial\rho_0/\partial \xi_+$ is the derivative w.r.t. $\xi_+$, which generates the necessary powers of the integration variable $p$ in \eqref{rhofinal}.

To calculate the modulus of the above expression, we decompose $\rho_0=\rho_1+i\rho_2$ and keep only terms up to quadratic in $\gamma_{1,2}(t)/\beta \sim g_0 k_BT/\hbar \Omega$ :
\begin{eqnarray}
    &&|\rho(x,x')|^2\propto 
    \left[1-8N\Delta^2\frac{\gamma^2_2(t)}{\beta^2}\right] |\rho_0|^2 -\nonumber \\
    &&\frac{4N\gamma_1(t)}{\beta}\partial|\rho_0|^2 +\frac{8N^2\gamma^2_1(t)}{\beta^2}\partial^2|\rho_0|^2+\nonumber\\
    &&16N\Delta\frac{\gamma_1(t)\gamma_2(t)}{\beta^2}\Im\left(\overline{\rho}_0\partial\rho_0\right)+\nonumber\\
    &&\frac{8N\gamma^2_1(t)}{\beta^2}\Re\left(\overline{\rho}_0\partial^2\rho_0\right)
\end{eqnarray}
We can now identify the (real) decoherence factor as the factor multiplying $|\rho_0(x,x')|^2$:
\begin{eqnarray}\label{Decfactorhightemperature}
    && |\Gamma_t(x-x')|^2=1-8N\Delta^2\frac{ \gamma^2_2}{\beta^2} \approx \exp\left[-8N\Delta^2\frac{ \gamma^2_2}{\beta^2}\right]\\
    && =\exp\left[-4N \frac {g^2_0 M} {\hbar^3\Omega\beta^2}|x-x'|^2\sin^2(\Omega t)\right]\\
    && =\exp\left[-2N \left(\frac{x-x'}{\lambda_{coh}}\right)^2\sin^2(\Omega t)\right]
\end{eqnarray}
Thus, the decay of spatial coherences is approximately Gaussian in the separation and controlled by the total amount of modes of both polarizations, $2N$, and the characteristic length given by:
\begin{align}\label{lcoh}
    \lambda_{coh} = \sqrt{\frac{\hbar^3\Omega\beta^2} {2g^2_0 M}} = \frac{ r}{\delta\varphi} \left(\frac{\hbar \Omega}{k_B T}\right),
\end{align}
where we in the last step used definition \eqref{g} of the coupling constant $g_0$. The coherence length is inversely proportional to the gravitational deflection angle $\delta\varphi$ and proportional to the average distance between the oscillator and the light $r$, rescaled by the ratio of the oscillator energy to the light (thermal) energy. Although the latter ration is small in the high temperature limit, the inverse of the deflection angle is orders of magnitude larger, making $\lambda_{coh}$ huge, unless the temperature is enormous. For example, using the parameters from \cite{Light_Matter_H} $r= 25 cm$, $M=10kg$, $\Omega= 2\pi\cdot 150 Hz$, we obtain for example micrometer-scale $\lambda_{coh}\approx 215 \mu m$ for $T=10^{20}K$. Using $N$ modes this enormous temperature can be scaled down by $1/\sqrt N$ but still remains huge for reasonable $N$.


\subsection{Single thermal mode}

For a simplified situation of a single thermal mode, i.e. $N=1$, \ref{Fczynnik} can be calculated analytically only assuming  $g_0 \ll 1$. We obtain:
\begin{eqnarray}
  &&  F_t(\Delta,p)\approx 1+2i\frac{\omega\left[\gamma_1(t)p+\gamma_2(t)\Delta\right]}{e^{\beta\omega}-1}\nonumber\\
    &&-\frac{2\omega^2(1+e^{\beta\omega})}{(e^{\beta\omega}-1)^2}\left[\gamma_1(t)p+\gamma_2(t)\Delta\right]^2
\end{eqnarray}
We can repeat the previous steps and obtain:
\begin{eqnarray}
    &&\rho(x,x')\propto\left[ 1 -\frac{2\omega^2\gamma^2_2\Delta^2}{(e^{\beta\omega}-1)^2}(1+e^{\beta\omega})+ 2i\frac{\omega\Delta\gamma_2(t)}{e^{\beta\omega}-1}\right]\rho_0\nonumber\\
    &&-\left[\frac{2\gamma_1(t)\omega }{e^{\beta\omega}-1} +4i \frac{\gamma_1(t)\gamma_2(t)\omega^2\Delta}{(e^{\beta\omega}-1)}+4i\frac{e^{\beta\omega}\gamma_1(t)\gamma_2(t)\omega^2\Delta}{(e^{\beta\omega}-1)^2}\right]\partial \rho_0 \nonumber\\
    &&- \left[\frac{2\gamma^2_1(t)\omega^2}{(e^{\beta\omega}-1)^2}+\frac{2\gamma^2_1(t)\omega^2e^{\beta\omega}}{(e^{\beta\omega}-1)^2}\right]\partial^2\rho_0
\end{eqnarray}
and the modulus:
\begin{eqnarray}
   && |\rho(x,x')|^2=\left[1-\frac{4\omega^2\Delta^2e^{\beta\omega}}{(e^{\beta\omega}-1)^2}\gamma^2_2(t)\right]|\rho_0|^2+ \nonumber\\
   &&\frac{2\omega\gamma_1(t)}{e^{\beta\omega}-1}\partial|\rho_0|^2+\frac{8\Delta\omega^2\gamma_1(t)\gamma_2(t)}{e^{\beta\omega}-1}\Im(\overline{\rho}_0\partial\rho_0)+\\
   &&\frac{2\omega^2\gamma^2_1(t)}{(e^{\beta\omega}-1)^2}\partial^2|\rho_0|^2+\frac{4\omega^2\gamma^2_1(t)e^{\beta\omega}}{(e^{\beta\omega}-1)^2}\Re(\overline{\rho}_0\partial^2\rho_0)\nonumber
\end{eqnarray}
from which we can read the decoherence factor:
\begin{eqnarray}
    &&|\Gamma_t(x-x')|^2= 1 - \frac{4\omega^2\Delta^2e^{\beta\omega}}{(e^{\beta\omega}-1)^2}\gamma^2_2(t)\\
    &&\approx \exp\left[-\frac{4\omega^2\Delta^2e^{\hbar\beta\omega}}{(e^{\hbar\beta\omega}-1)^2}\gamma^2_2(t)\right]\\
    && = \exp\left[-\frac{2g^2_0M}{\hbar\Omega}\frac{\omega^2e^{\hbar\beta\omega}}{(e^{\hbar\beta\omega}-1)^2}
    |x-x'|^2\sin^2(\Omega t)\right]\\
    && = \exp\left[-\left(\frac{x-x'}{\lambda_{coh}}\right)^2\sin^2(\Omega t)\right].
\end{eqnarray}
It is easy to see that for $\beta\omega \ll 1$ it becomes equal to (\ref{Decfactorhightemperature}). The coherence length is now given by:
\begin{eqnarray}
    \lambda_{coh}=\sqrt{\frac{\hbar\Omega}{2g^2_0M}}\frac{e^{\hbar\beta\omega}-1}{\omega e^{\hbar\beta\omega/2}}=\frac{r}{\delta\phi}\left(\frac{\hbar\Omega}{\overline{E}}\right)\sqrt{\frac{\overline{n}}{1+\overline{n}}},
\end{eqnarray}
where $\overline E=\hbar\omega/(e^{\hbar\beta\omega}-1)$ is the light's thermal energy and $\overline n = 1/(e^{\hbar\beta\omega}-1)$ is the mean photon number. We see a similar structure to \eqref{lcoh} with the modification of the thermal energy part.


\section{Decoherence from coherent light }

In this section we study a simpler example, when the oscillator is initially in the ground state and the light is a single-mode ($N=1$) coherent state: 
\begin{eqnarray}
    \rho_{SE}(0) =\ket{0}\bra{0} \otimes \ket{\alpha}\bra{\alpha},
\end{eqnarray}
It is easy to find the matrix elements \eqref{generalform}  in this case due to the simple form of the characteristic function of the oscillator: $\chi(\eta) = e^{-|\eta|^2/2}$ and the probability $p(E)$, which in this case is simple given by the elements of the Fock-basis expansion $p(n)=e^{-|\alpha|^2/2}\sum_{n=0}^{\infty}\alpha^n/n!$. Substituting those into (\ref{generalform}), we obtain:
    \begin{eqnarray}
    &&\rho(x,x') = \sqrt{\frac{\kappa}{\pi^2}}e^{-\frac{\Delta^2}{2}-|\alpha|^2}\int dp e^{-\frac{p^2}{2}}e^{-i p\xi_+}\nonumber \\
    &&\times \sum^{\infty}_{n=0} \frac{1}{n!}\left[|\alpha|^2 e^{2i\omega(\gamma_1(t) p + \gamma_2(t)\Delta)}\right]^n\\
   && =\sqrt{\frac{\kappa}{\pi^2}}e^{-\frac{\Delta^2}{2}} \int dp \exp\left(-\frac{p^2}{2}-i p\xi_+\right)\\
    &&\times \exp\left\{-|\alpha|^2\left[1-e^{2i\omega(\gamma_1(t) p +\gamma_2(t)\Delta)}\right]\right\}\nonumber,
\end{eqnarray}
where the last line is the expression for the influence function $F_t(p;\Delta)$, cf. \eqref{rhofinal}, \eqref{F}. To calculate the above integral, we will use again the fact that 
$g_0 \ll 1$  and  expand the $\gamma$-dependent exponential up to $\mathcal{O}(g^2_0)$, since, as before, we will be calculating the modulus squared of the matrix elements. We obtain:
\begin{eqnarray}
    &&|\rho(x,x')|^2\approx \frac{\kappa}{\pi\sigma(t)}\exp\left[-\Delta^2-\frac{2\xi^2_+}{\sigma(t)}-4|\alpha|^2\omega^2\gamma^2_2(t)\Delta^2\right]\nonumber\\
    &&\exp\left[-2|\alpha|^4\gamma^2_1(t)\omega^2/\sigma(t)- 4|\alpha|^2\gamma_1(t)\xi_+\omega/\sigma(t)\right],
\end{eqnarray}
where $\sigma(t)=\frac{1}{2}-2|\alpha|^2\omega^2\gamma^2_1(t)$. As before, we identify the decoherence factor as the factor that depends both on the separation $\Delta$ and the coupling to the environment $g_0$: 
\begin{eqnarray}
&&|\Gamma_t(x-x')|^2=\exp\left[-4|\alpha|^2\omega^2\gamma^2_2(t)\Delta^2\right]\\
&& = \exp\left[-\frac{2g^2_0M\omega^2}{\hbar\Omega}|\alpha|^2|x-x'|^2\sin^2(\Omega t)\right]\\
&&=\exp\left[-\left(\frac{x-x'}{\lambda_{coh}}\right)^2\sin^2(\Omega t)\right],
\end{eqnarray}
where the coherence length is identified as:
\begin{eqnarray}
    \lambda_{coh}=\sqrt{\frac{\hbar \Omega}{2g^2_0\omega^2M |\alpha|^2}}=\frac{r}{|\alpha|\delta\phi}\left(\frac{\Omega}{\omega}\right),
\end{eqnarray}
and is inversely proportional to the strength of the coherent light $|\alpha|$.

\section{Conclusions and final remarks}
We studied potential decoherence effects due to the gravitational light bending. We considered a massive harmonic oscillator interacting with quantized light at a fixed distance via classical light bending interaction of general relativity, proposed in \cite{Light_Matter_H}. Assuming the light is unobserved and hence can be traced over, 
we derive decoherence lengthscales in a fairly general scenario where the state of the oscillator is arbitrary and the light is either thermal or coherent light. We use the fact that the gravitational interaction is very weak on the mass scales where quantum effects could still be imaginable. Our lengthscales depend, among other parameters, on the classical gravitational deflection angle. As one can imagine, this angle is extremely small for non-astronomical bodies, requiring enormously energetic light in order for the decoherence effects to manifest  themselves on real-life, microscopic scales. 

Despite that, we show that if the postulated coupling scheme works, gravity mediated interaction will lead to decoherence effects. The effect is somewhat similar to the proposed universal gravitational decoherence \cite{Pikovski2015}. It relies on a postulated quantum-mechanical coupling between the gravitational energy in an external gravitational field of the centre of mass of a compound system and its internal energy. Here the situation is different -- although the central oscillator and the light can be treated as a compound system with the light playing the role of internal degrees of freedom, but there is no external gravitational field. It is thus an interesting question if the scheme of the universal gravitational decoherence can be generalized to the situation of a compound system with constituents interacting via the gravitational field. The works on entanglement in the linearized gravity \cite{Christodoulou2023} may give a hint here. 

Another remark is that the initial product state between the mass and the light, assumed in deriving \eqref{rhoSE}, while realistic in optomechanics, here is rather artificial as one cannot switch off the gravitational field of a mass. The situation resembles the well-known charge-field separation issue in the basic, non-relativistic quantum electrodynamics, which leads to the effective mass renormalization (dressing).  It is an interesting question if one can define a gravitational analogue of the mass dressing as well. 

Finally, let us remark that for the thermal light, the decoherence process is not accompanied by the information transfer to the environment. Indeed, the interaction Hamiltonian \eqref{Htot} commutes with any initial state of the environment, diagonal in the energy basis. Another way to see it, is by calculating the state of the environment from \eqref{rhoSE}: $\rho_E(t)=\tr_S\rho_{SE}(t)=\sum_E \chi(0,0) \bra{E}\rho_{E}\ket{E}\ket{E}\bra{E} = \rho_{E}$. Thus, although there is a decoherence effect, there is no quantum Darwinism \cite{Zurek2009}, \cite{Korbicz2021} happening in this case. This is a general feature of the gravitational coupling, which couples to the energy and in order to have Darwinism, one has to have some initial coherences in the energy basis \cite{Korbicz2017} (see also \cite{Tuziemski2019}).

\section*{Acknowledgements}
TB and JKK acknowledge the financial support of the National Science Centre (NCN) through the QuantEra project Qucabose 2023/05/Y/ST2/00139.

\begin{figure}[t]
    \centering
    \includegraphics[width=0.35\textwidth]{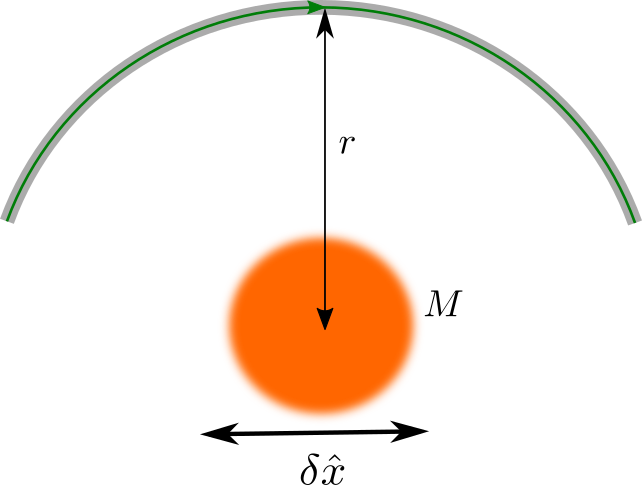} 
    \caption{The setup: the mass is harmonically trapped while the light is guided in a half-circular cavity at a (macroscopically) constant distance from the mass. The light bending happens within the cavity.}
    \label{fig}
\end{figure}

\appendix
\section*{Appendix}

We briefly recall here the origin of the interaction Hamiltonian \eqref{V}. The main object is the classical effective potential, experienced by the light: 
\begin{equation}\label{Vclass}
    V=-\frac{2GM\omega_k}{r}\varepsilon^*_{k'\nu'}\cdot \varepsilon_{k\nu},
\end{equation}
where $\varepsilon_{k\nu}$, $\varepsilon_{k'\nu'}$ are the polarization vectors of the incoming and outgoing light respectively and $\omega_k$ its frequency. An interesting way to derive it is via a hypothesis that the linearized gravitational field can be quantized using the standard field quantization approach, as it was done in  \cite{Scadron1979} and then in more detail in \cite{Light_Matter_H}. Considering a tree-level photon-matter scattering process, intermediated via a (hypothetical) graviton, one can obtain \eqref{Vclass} via the standard S-matrix methods, with some additional technical assumptions, e.g. of a small scattering angle $(k-k')^2\approx 0$ and a static limit  as the mass is assumed to be large enough not to feel the decoy of the photons \cite{PhysRevD.105.106028}; we refer the reader to the detailed derivation presented the Appendix A of \cite{Light_Matter_H}. Changing to the operator formalism, the potential \eqref{Vclass} should be treated as a quantum operator, see e.g. \cite{PhysRevD.105.106028}. However, one immediately encounters the problem with a non-local operator $1/\hat r$. A great simplification, enough for this type of proof-of-principle considerations, comes if one fixes the distance $r$ as "macroscopic variable", e.g. in a setup shown in Fig.~\ref{fig} \cite{Light_Matter_H}. Assuming that the mass can perform small oscillations $\delta x$, the mass-to-cavity distance reads:
\begin{equation}
    \vec{r}=\begin{pmatrix}
    \delta x- r\cos\theta\\
    - r\sin\theta\\
    0
\end{pmatrix},
\end{equation}
which leads to the expansion of \eqref{Vclass}: 
\begin{equation}\label{Vexp}
    V\approx -2MG\omega_k\left(\frac{1}{r}+\frac{2\delta x}{r^2}\cos\theta+\mathcal{O}(\delta x^2)\right)\varepsilon^*_{k'\nu'}\cdot \varepsilon_{k\nu}.
\end{equation}
The first term in \eqref{Vexp} is just a constant energy shift and can be omitted, while the angular coordinate $\theta$ can be integrated over the path of light ray to give an angle averaged potential. Now, the small oscillation $\delta x$ is quantized and promoted to an operator $\delta x \to \delta \hat x$, which leads to a quantum potential: 
\begin{equation}
    \hat V=  -\frac{4MG\omega_k}{r^2}\delta \hat x \otimes \hat \varepsilon^*_{k'\nu'}\cdot \hat \varepsilon_{k\nu}, 
\end{equation}
where the polarization vectors have been promoted to quantum operators as well: $\varepsilon_{k\nu} \to \hat \varepsilon_{k\nu} = \varepsilon_{k\nu} \hat a_{k\nu}$. Further simplifying the problem by assuming $k'\approx k,  \ \nu'\approx\nu$, and assuming harmonic oscillations $\delta \hat{x} =(\hat{b}^{\dagger}+\hat{b})/\sqrt{2M\Omega}$, finally leads to \eqref{V}:
\begin{eqnarray}
    \hat V_{k\nu}=-g_0(\hat b^\dagger+\hat b)\otimes \hbar \omega_k \hat a^{\dagger}_{k\nu} \hat a_{k\nu}
\end{eqnarray}
We then omit the hats for the notational simplicity.

\bibliography{bibliografia}
\end{document}